\documentclass{article}
\usepackage{amsfonts}
\usepackage{amsmath}

\input{tcilatex}

\begin{document}

\title{Protein Folding as a Physical Stochastic Process\thanks{%
Invited talk at BIOMAT 2007 International Symposium on Mathematical and
Computational Biology, Zfmzcao dos Buzios, RJ, Brazil, 24th-29th Nov, 2007}}
\author{Kerson Huang \\
Physics Department, Massachusetts Institute of Technology\\
Cambridge, MA, USA 02139}
\maketitle

\begin{abstract}
We model protein folding as a physical stochastic process as follows. The
unfolded protein chain is treated as a random coil described by SAW
(self-avoiding walk). Folding is induced by hydrophobic forces and other
interactions, such as hydrogen bonding, which can be taken into account by
imposing conditions on SAW. The resulting model is termed CSAW (conditioned
self-avoiding walk. Conceptually, the mathematical basis is a generalized
Langevin equation. In practice, the model is implemented on a computer by
combining SAW and Monte Carlo. To illustrate the flexibility and
capabilities of the model, we consider a number of examples, including
folding pathways, elastic properties, helix formation, and collective modes.
\end{abstract}

\section{Introduction}

One of the outstanding unsolved problems in molecular biology is protein
folding\cite{bra} \cite{dag}. The principle through which the amino acid
sequence determines the native structure, as wells as the dynamics of the
process, remain open questions. Generally speaking, there have been two
types of approaches to the problem: bioinformatics\cite{kan} and molecular
dynamics (MD)\cite{MD}.

Bioinformatics is purely data analysis, and does not involve dynamics at
all. It massages the data base of known proteins in different ways, using
very sophisticated computer programs, in order to discover correlations
between sequence and structure. By its very nature, it cannot provide any
physical understanding.

On the other hand, MD solves the Newtonian equations of motion of all the
atoms in the protein on a computer, using appropriate inter-atomic
potentials. To describe the solvent, one inculdes thousands of water
molecules explicitly, treating all the atoms in the water on same footing as
those on the protein chain. Not surprisingly, such an extravagant use of
computing power is so inefficient that one can follow the folding process
only to about a microsecond, whereas the folding of a real protein takes
from one second to ten minutes.

We shall try an approach from the point of view of statistical mechanics\cite%
{hualec}. After all, the protein is a chain molecule immersed in water, and,
like all physical systems, will tend towards thermodynamic equilibrium with
the environment. Our goal is to design a model that embodies physical
principles, and at the same time amenable to computer simulation in
reasonable time.

We treat the protein as a chain performing Brownian motion in water,
regarded as a medium exerting random forces on the chain, with the
concomitant energy dissipation. In addition, we include regular (non-random)
interactions within the chain, as well as between the chain and the medium.

The unfolded chain is assumed to be a random coil described by SAW
(self-avoiding walk), as suggested by Flory\cite{flo} some time ago. That
is, each link in the chain corresponds to successive random walks, in which
the chain is prohibited from revisiting an occupied position. Two types of
interactions are included in our initial formulation:

\begin{itemize}
\item the hydrophobic action due to the medium, which causes the chain to
fold;

\item the hydrogen-bonding within the chain, which leads to helical
structure.
\end{itemize}

\noindent Other interactions can be added later.

We model the protein chain in 3D space, keeping only degrees of freedom
relevant to folding, which we take to be the torsional angles between
successive links. In the computer simulation, we first generate an ensemble
of SAW's, and then choose a subensemble through a Monte Carlo method, which
generates a canonical ensemble with respect to a Hamiltonian that specifies
the interactions. We call the model CSAW\cite{hua0} \cite{hua1} (conditioned
self-avoiding walk). Mathematically speaking, it is based on a Langevin
equation\cite{hualec} describing the Brownian motion of a chain with
interaction. There seems little doubt that such an equation does describe a
protein molecule in water, for It is just Newton's equation with the
environment treated as a stochastic medium.. The model can be implemented
efficiently on a computer, and is flexible enough to be used as a
theoretical laboratory.

Both CSAW and MD are based on Newtonian mechanics, and differ only in the
idealization of the system. In CSAW we replace the thousands of water
molecules used in MD by a stochastic medium --- the heat reservoir of
statistical mechanics. We ignore inessential degrees of freedom, such as
small fluctuations in the lengths and angles of the chemical bonds that link
the protein chain. The advantages of these idealizations are that

\begin{itemize}
\item we avoid squandering computer power on irrelevant calculations;

\item we gain a better physical understanding of the folding process.
\end{itemize}

One often hears a debate on whether the folding process is "thermodynamic"
or "kinetic". There is also an oft cited \textquotedblleft Levinthal
paradox", to the effect that the folding time should be much larger than the
age of the universe, since the protein (presumably) had to search through an
astronomically large number of states before finding the right one. From our
point of view, these are not real issues.

The question of thermal equilibrium merely hangs on whether the protein can
reach equilibrium in realistic time, instead being trapped in some
intermediate state. For any particular protein, simulation of the Langevin
equation will answer the question.

As to Levinthal's "paradox", the protein is blithely unaware of that. It
just follows pathways guided by the Langevin equation.

After a brief review of the basics of protein folding and stochastic
processes, we shall describe the model in more detail, and illustrate its
use through examples involving realistic protein fragments. We will
demonstrate folding pathways, elastic properties, helix formation, and
protein collective modes.

The results indicate that the model has been successful in describing
qualitative features of folding in simple proteins.

\section{Protein basics}

\subsection{ The protein chain}

The protein chain consists of a sequence of units or "residues", which are
amino acids chosen from a pool of 20. This sequence is called the \textit{%
primary structure.} The center of each amino acid is a carbon atom called $%
C_{\alpha }$. Along the protein chain, the $C_{\alpha }$'s are connected by
covalent chemical bonds in the shape of a \textquotedblleft crank" that lies
in one plane. Two cranks join at a $C_{\alpha }$ with a fixed angle between
them, the tetrahedral angle $\theta _{\text{tet}}=-\arccos \left( 1/3\right)
\approx 110^{\circ }$. The amino acids differ from each other only in the
side chains connect to the $C_{\alpha }$'s. There are 20 possible choices
for side chains.

The relative orientation of successive cranks is determined by two torsional
angles $\phi $ and $\psi $, as schematically illustrated in Fig.1. These
torsional angles are the only degrees of freedom relevant to protein
folding, and small oscillations in bond lengths and bond angles can be
ignored. For our purpose, therefore, a protein of $N$ residues has $2(N-1)$
degrees of freedom.\FRAME{ftbpFU}{3.2846in}{1.2808in}{0pt}{\Qcb{Schematic
representation of the protein chain. Centers of residues are carbon atoms
labeled $\protect\alpha $. They are connected by rigid chemical bonds in the
shape of a planar crank. The only degrees of freedom we consider are the
torsional angles $\protect\phi ,\protect\psi $ that specify the relative
orientations of successive cranks. Residues can differ only in the side
chains labeled $R_{i},$ chosen from a pool of twenty. Atoms connected to the
cranks are omitted for clarity.}}{}{schematicchain.jpg}{\special{language
"Scientific Word";type "GRAPHIC";maintain-aspect-ratio TRUE;display
"USEDEF";valid_file "F";width 3.2846in;height 1.2808in;depth
0pt;original-width 6.7101in;original-height 2.5996in;cropleft "0";croptop
"1";cropright "1";cropbottom "0";filename
'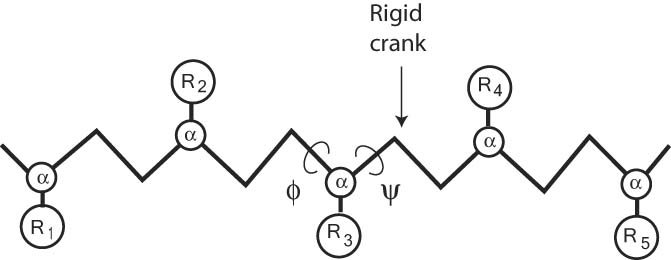';file-properties "XNPEU";}}

\subsection{Secondary and tertiary structures}

At high temperatures, or in an acidic solution, the protein exists in an
unfolded state that can be represented by a random coil\cite{flo}. When the
temperature is lowered, or when the solution becomes aqueous, it folds into
a "native state" of definite shape. Fig.2 shows the native state of
myoglobin with different levels of detail. Local structures, such as
helices, are called \textit{secondary structures}. When these are blurred
over, one sees a skeleton called the \textit{tertiary structure}.\FRAME{%
ftbpFU}{3.525in}{1.657in}{0pt}{\Qcb{Native state of Myoglobin showing
different degrees of detail.}}{}{myoglobin.jpg}{\special{language
"Scientific Word";type "GRAPHIC";maintain-aspect-ratio TRUE;display
"USEDEF";valid_file "F";width 3.525in;height 1.657in;depth
0pt;original-width 7.8205in;original-height 3.6599in;cropleft "0";croptop
"1";cropright "1";cropbottom "0";filename
'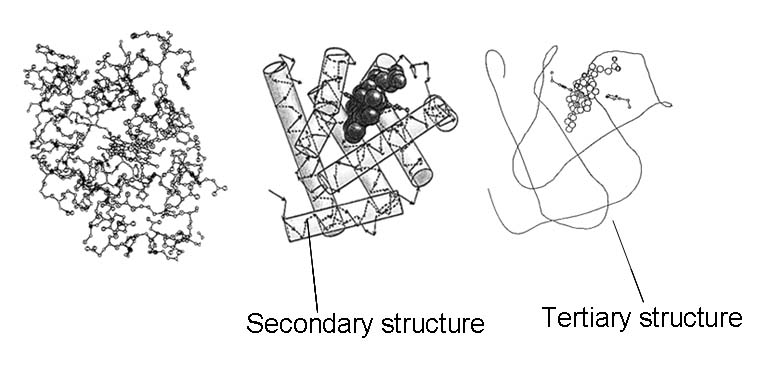';file-properties "XNPEU";}}

Secondary structures are of two main types, the alpha helix and the beta
sheet, as shown in Fig.3. The former is stabilized by hydrogens bonds that
connect \ residues 1 to 4, 2 to 5, \textit{etc}. The beta sheet is a global
mat sewn together by hydrogen bonds.\FRAME{ftbpFU}{3.2707in}{2.4094in}{0pt}{%
\Qcb{Secondary structures. Dotted lines in the alpha helix denote hydrogen
bonds. The beta sheet is composed of \textquotedblleft beta strands" matted
together by hydrogen bonds. two ajacent strands are connect by a
\textquotedblleft beta hairpin".}}{}{alphabeta.jpg}{\special{language
"Scientific Word";type "GRAPHIC";display "USEDEF";valid_file "F";width
3.2707in;height 2.4094in;depth 0pt;original-width 4.2194in;original-height
3.4402in;cropleft "0";croptop "1";cropright "1";cropbottom "0";filename
'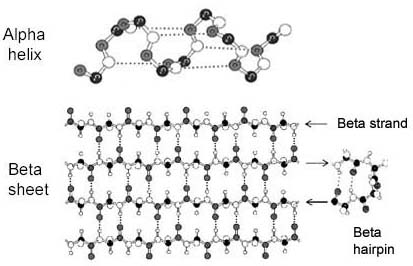';file-properties "XNPEU";}}

\subsection{Hydrophobic effect}

The molecules of liquid water form hydrogen bonds with each other, resulting
in a dense fluctuating network, in which bonding partners change on a time
scale of 10$^{-12}$s. A computer simulation of such a network is shown in
Fig.4a\cite{water}. A foreign molecule introduced into water disrupts the
network, unless it can participate in hydrogen bonding. If it can
hydrogen-bond with water, it is said to be \textquotedblleft soluble", or
\textquotedblleft hydrophilic", and will be received by water molecules as
one of their kind. Otherwise it is unwelcome, and said to be
\textquotedblleft insoluble", or "hydrophobic". Protein side chains can be
hydrophilic or hydrophobic.

When immerse in water, the protein chain folds in order to shield the
hydrophobic residues from water. In effect, the water network squeezes the
protein into shape. This is called the "hydrophobic effect". However, a
\textquotedblleft frustration" arises in this process, because the skeleton
is hydrophilic, and likes to be in contact with water, as indicated in
Fig.4b. The frustration is resolved by the formation of secondary
structures, which use up hydrogen bonds internally. The folded chain reverts
to a random coil when the temperature becomes too high, or when the pH of
the solution becomes acidic.\FRAME{ftbpFU}{4.3535in}{1.6717in}{0pt}{\Qcb{(a)
Computer simulation of network of hydrogen bonds in liquid water. (b) The
hydrophobic side chains $R_{1}$ and $R_{2}$ cannot form hydrogen bonds, and
prefer to be shielded from water. However, the atoms $O$ and $H$ on the main
chain need to form hydrogen bonds. A \textquotedblleft frustration" thereby
arises, and is resolved by formation of secondary structures that use up\
hydrogen bonds internally. }}{}{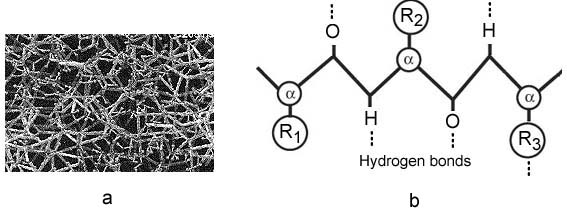}{\special{language "Scientific
Word";type "GRAPHIC";maintain-aspect-ratio TRUE;display "USEDEF";valid_file
"F";width 4.3535in;height 1.6717in;depth 0pt;original-width
5.6697in;original-height 2.1594in;cropleft "0";croptop "1";cropright
"1";cropbottom "0";filename 'figs/water.jpg';file-properties "XNPEU";}}

\subsection{Folding stages}

As depicted schematically in Fig.5, a typical folding process consists of a
very rapid collapse into an intermediate state called the \textquotedblleft
molten globule". The latter takes a relatively long time to undergo fine
adjustments to reach the native state. The collapse time is generally less
than 200 $\mu $s, while the molten globule can last as long as 10 minutes.%
\FRAME{ftbpFU}{4.7106in}{1.8775in}{0pt}{\Qcb{Being squeezed by a water net,
the protein chain rapidly collapses into the molten globule state, which
slowly adjusts itself into the native state.}}{}{foldingstages.jpg}{\special%
{language "Scientific Word";type "GRAPHIC";maintain-aspect-ratio
TRUE;display "USEDEF";valid_file "F";width 4.7106in;height 1.8775in;depth
0pt;original-width 7.8698in;original-height 3.1194in;cropleft "0";croptop
"1";cropright "1";cropbottom "0";filename
'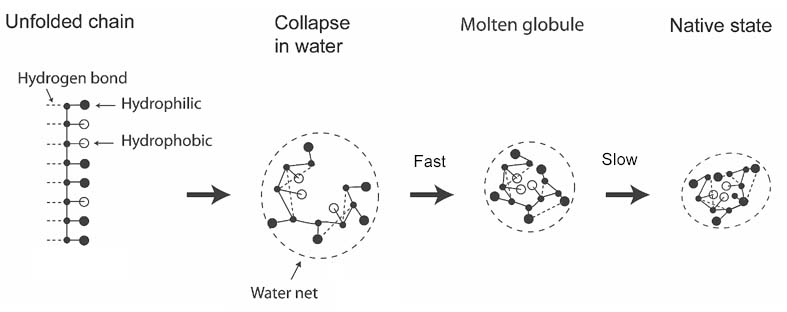';file-properties "XNPEU";}}

\subsection{Statistical nature of the folding process}

We have to distinguish between protein assembly inside a living cell (%
\textit{in vivo}), and folding in a test tube (\textit{in vitro}). In the
former the process takes place within factory molecules called ribosomes,
and need the assistance of "chaperon" molecules to prevent premature
folding. In the latter, the molecules freely fold or unfold, reversibly,
depending on the pH and the temperature.

We deal only with folding \textit{in vitro}, in which ten of thousands of
protein molecules undergo the folding process independently, and they do not
fold in unison. We are thus dealing with an ensemble of protein molecules,
in which definite fractions exist in various stages of folding at any given
time. The Langevin equation naturally describes the time evolution of such
an ensemble. Behavior of individual molecules fluctuate from the average,
even after the ensemble has reached equilibrium. In macroscopic systems
containing the order of 10$^{23}$ atoms, such fluctuations are unobservably
small. For a protein with no more than a few thousand atoms, however, these
fluctuations are expected to be pronounced.

\section{Stochastic process}

\subsection{Stochastic variable}

A stochastic process is one involving random forces, and is described
through a so-called stochastic variable (or random variable), which does not
have a definite value, but is characterized instead by a probability
distribution of values. Practically everything we deal with in the
macroscopic world involve random variables, from the position of a billiard
ball to the value of a stock.

Einstein pointed out the essence of a stochastic variable in his theory of
Brownian motion. He emphasized that every Brownian step we can observe is
the result of a very large number of smaller random steps, which in turn are
the result of a very number of even smaller steps, and so on, until we reach
the cutoff imposed by atomic structure. This self-similarity leads to a
Gaussian distribution, regardless of the underlying mechanism --- a result
known as the \textit{central limit theorem}\cite{huastat}.

\subsection{Brownian motion}

The simplest stochastic process is the Brownian motion of a single particle
suspended in a medium. Its position $x(t)$ is a stochastic variable
described by the Langevin equation 
\begin{equation}
m\ddot{x}=F(t)-\gamma \dot{x}
\end{equation}%
Here, the force exerted by the medium on the particle is split into two
parts: a randomly fluctuating force $F\left( t\right) $ and a friction $%
-\gamma \dot{x}$. The random force is a member of a statistical ensemble
with the properties%
\begin{align}
\left\langle F(t)\right\rangle & =0  \notag \\
\left\langle F(t)F(t^{\prime })\right\rangle & =c_{0}\delta \left(
t-t^{\prime }\right)
\end{align}%
where the brackets $\langle \rangle $ denote ensemble average. The two
forces are not independent, but related through the fluctuation-dissipation
theorem:%
\begin{equation}
\frac{c_{0}}{2\gamma }=k_{B}T
\end{equation}%
where $k_{B}$ is Boltzmann's constant and $T$ is the absolute temperature, a
property of the medium.

The Langevin equation can be solved exactly, and also be simulated by 
\textit{random walk}. Both methods lead to diffusion, in which the position
has a Gaussian distribution with variance $\sqrt{2Dt}$, where $t$ is the
time, and $D=c_{0}/\left( 2\gamma ^{2}\right) $ is called the diffusion
constant. An equivalent expression is \textit{Einstein's relation}%
\begin{equation}
D=\frac{k_{B}T}{\gamma }
\end{equation}%
Thus, a random force must generate energy dissipation, and the dissipation
constant $\gamma $ can be deduced from the variance of the distribution of
positions.

\subsection{Monte Carlo}

If a particle undergoes Brownian motion in the presence of a regular
(non-random) external force $G(x)$, we may not be able to solve the Langevin
equation exactly, but we can still simulate it on a computer by \textit{%
conditioned random walk,} as follows\textit{. }We first generate a random
trial step, but accept it only according to the Monte Carlo algorithm. Let $%
E $ be the potential energy corresponding to the external force $G$. Let $%
\Delta E$ be the energy change in the proposed update. The algorithm is as
follows:

\begin{itemize}
\item if $\Delta E\leq 0,$ accept it;

\item if $\Delta E>0,$ accept it with probability $\exp \left( -\Delta
E/k_{B}T\right) .$
\end{itemize}

\noindent The last condition simulates thermal fluctuations, which may drive
the system to a higher energy. After a sufficiently large number of updates,
the sequence of state generated will yield a canonical ensemble with
temperature $T$ . That is, the Monte Carlo procedure tends to minimize not
the energy, but the free energy.

Mathematically speaking, conditioned random walk simulates a generalized
Langevin equation, as indicated in the following: 
\begin{equation}
m\ddot{x}=\underset{\text{Treat via random walk}}{\left[ F(t)-\gamma \dot{x}%
\right] }+\underset{\text{Treat via Monte Carlo}}{G(x).}
\end{equation}%
Of course, we could integrate the whole equation as a stochastic
differential equation, as an alternative to Monte Carlo. The equivalence of
these two methods is illustrated by example in the appendix of Ref.\cite{lei}%
.

\section{CSAW}

In protein folding, we are dealing with the Brownian motion of a chain with
interactions. All we need to do, in principle, is to generalize conditioned
random walk to conditioned SAW (self-avoiding walk). The resulting model is
called CSAW (conditioned self-avoiding walk).

We can generate a SAW representing an unfolded protein chain by the \textit{%
pivot algorithm}\cite{li} \cite{ken}, as follows. Choose an initial chain in
3D continuous space, and hold one end of the chain fixed.

\begin{itemize}
\item Choose an arbitrary point on the chain as pivot.

\item Rotate the end portion of the chain rigidly about the pivot (by
changing the torsional angles at the pivot point).

\item If this does not result in any overlap, accept the configuration,
otherwise repeat the procedure.
\end{itemize}

\noindent By this method, we can generate a uniform ergodic ensemble of
SAW's, which simulates a Langevin equation of the form

\begin{equation}
m_{k}\mathbf{\ddot{x}}_{k}=\mathbf{F}_{k}(t)-\gamma _{k}\mathbf{\dot{x}+U}%
_{k}\text{,\ \ \ \ (}k=1,\cdots ,N)\text{\ \ \ \ \ \ \ \ \ \ \ }
\end{equation}%
where the subscripts $k$\ label the residues along the chain. The terms $%
\mathbf{U}_{k}$ denote the regular (non-random) forces that maintain the
rigid bonds between successive residues, and that prohibit the residues from
overlapping one another.

We now add other regular forces $G_{k},$ which include the hydrophobic
interaction and hydrogen-bonding. Treating this force via Monte Carlo
results in CSAW, which simulates a generalized Langevin equation as
indicated in the following:

\begin{equation}
m_{k}\mathbf{\ddot{x}}_{k}=\,\underset{\text{ \ \ Treat via SAW}}{\left( 
\mathbf{F}_{k}-\gamma _{k}\mathbf{\dot{x}+U}_{k}\right) }+\underset{\text{%
Treat via Monte-Carlo}}{\mathbf{G}_{k}.}\text{ \ \ (}k=1\cdots N)
\end{equation}%
Now we shall specify the forces $G_{k}$ explicitly.

\section{Implementation of CSAW}

To reiterate, the system under consideration is a sequence of centers
corresponding to $C_{\alpha }$ atoms, connected by planar \textquotedblleft
cranks". The degrees of freedom of the system are the pairs of torsional
angles $\left\{ \phi _{i},\psi _{i}\right\} $ specifying the relative
orientation of two successive cranks. There are $O$ and $H$ atoms attached
to each crank, through rigid bonds lying in the same plane as the crank. The
residues can differ from one another only through the side chains attach to $%
C_{\alpha }$, and there are 20 of them to choose from. As indicated in
Fig.6, the center of the side chain is located at an apex of a tetrahedron
with $C_{\alpha }$ at the center.\FRAME{ftbpFU}{2.5685in}{2.9343in}{0pt}{%
\Qcb{ The side chain is at the apex of a tetrahedron with $C_{\protect\alpha %
}$ at the center.}}{}{chaindetails.jpg}{\special{language "Scientific
Word";type "GRAPHIC";maintain-aspect-ratio TRUE;display "USEDEF";valid_file
"F";width 2.5685in;height 2.9343in;depth 0pt;original-width
4.9in;original-height 5.5997in;cropleft "0";croptop "1";cropright
"1";cropbottom "0";filename '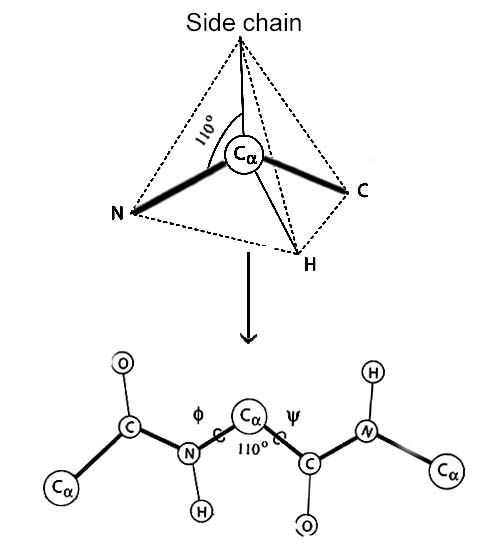';file-properties "XNPEU";}%
}

We can start with a chain of bare cranks, and then add other components one
by one, as desired. We can first represent the side chains by hard spheres,
and put in the atoms in a more elaborate version. In this manner, we can
tinker with different degrees of buildup, and investigate the relative
importance of each element.

For Monte Carlo, we take the energy $E$ to be

\begin{align}
E& =-g_{1}K_{1}-g_{2}K_{2}  \label{energy} \\
K_{1}& =\text{Total contact number of all hydrophobic residues}  \notag \\
K_{2}& =\text{Number of hydrogen bonds}  \notag
\end{align}

The first term in $E$ expresses the hydrophobic effect. The contact number
of a residue is the number of atoms touching its side chain. In the simplest
version, in which we do not explicitly put in the side chain, the contact
number is simply the number of atoms in contact with $C_{\alpha }$, \textit{%
not counting the other }$C_{\alpha }$'s\textit{\ lying next to it along the
chain}. This is illustrated in Fig.7a.

\FRAME{ftbpFU}{3.6754in}{2.437in}{0pt}{\Qcb{(a) The shaded hydrophobic
residue illustrated here has four contact neighbors. The permanent neighbors
along the chain are not counted. (b) Hydrogen bonding occurs between $O$ and 
$H$ on the main chain, from different residues. }}{}{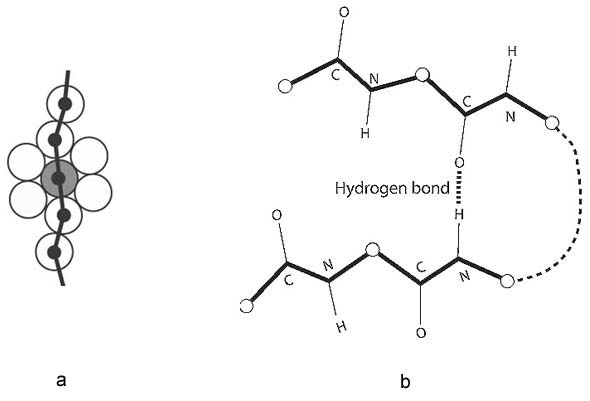}{%
\special{language "Scientific Word";type "GRAPHIC";maintain-aspect-ratio
TRUE;display "USEDEF";valid_file "F";width 3.6754in;height 2.437in;depth
0pt;original-width 6.5596in;original-height 4.3405in;cropleft "0";croptop
"1";cropright "1";cropbottom "0";filename
'figs/interactions.jpg';file-properties "XNPEU";}}

The contact number measures how well a residues is being shielded from the
medium. When two hydrophobic residues are in contact, the total contact
number increases by 2, an this induces an effective attraction between
hydrophobic residues. The unfolded chain corresponds to $g_{1}=0.$

The second term in $E$ describes hydrogen bonding. As illustrated in Fig.7b,
a hydrogen bond is deemed to have formed between $O$ and $H$ $\ $from
different cranks when

\begin{itemize}
\item the distance between $O$ and $H$ is 2.5 A, within given tolerance;

\item The bonds $C=O$ and $N-H$ are antiparallel, within given tolerance.
\end{itemize}

Only the combinations $g_{1}/k_{B}T$ and $g_{2}/k_{B}T$ appear in the Monte
Carlo procedure. They are treated as adjustable parameters.

Note that $E$ only includes the potential energy. We can leave out the
kinetic energy because it contributes only a constant factor to the
configurational probability of the ensemble.

\ 

\section{Exploratory runs}

It is instructive to run the program with minimal components, as described
in Refs.\cite{hua0} \cite{hua1}. For a chain of 30 residues, the main
findings are the following:

\begin{itemize}
\item Under hydrophobic forces alone, without hydrogen-bonding, the chain
folds into a reproducible shape. This shows that the hydrophobic effect
alone can produce tertiary structure. There is no secondary structure in
this case, and the chain rapidly collapses to the final structure without
passing through an intermediate state..

\item When there is no hydrophobic force and the interaction consists purely
of hydrogen-bonding, the chain rapidly folds into one long alpha helix.

\item When both hydrophobic force and hydrogen bonding are taken into
account, secondary structure emerges. The folding process exhibits two-stage
behavior, with a fast collapse followed by slow \textquotedblleft
annealing", in qualitative agreement with experiments.
\end{itemize}

We now recount some simulations of realistic protein fragments.

\section{Folding pathways and energy landscape}

Chignolin is a synthetic peptide of 10 residues \cite{chignolin}, in the
shape of a \textquotedblleft beta hairpin" -- a turn in a beta sheet as
depicted in Fig.3. Jinzhi Lei \cite{lei0} of Tsinghua University modeled it
in CSAW, with side chains modeled as hard spheres. The native state emerges
after about 70000 trial steps, as shown in Fig.8. The computation took less
than 5 minutes on a work station. In contrast, an MD simulation on the same
work station did not reach the native state in one month's computation. The
run was repeated 100 times independently, to obtain an ensemble of folding
paths.\FRAME{ftbpFU}{4.3846in}{2.6757in}{0pt}{\Qcb{Folding of Chignolin, a
beta hairpin with ten residues.}}{}{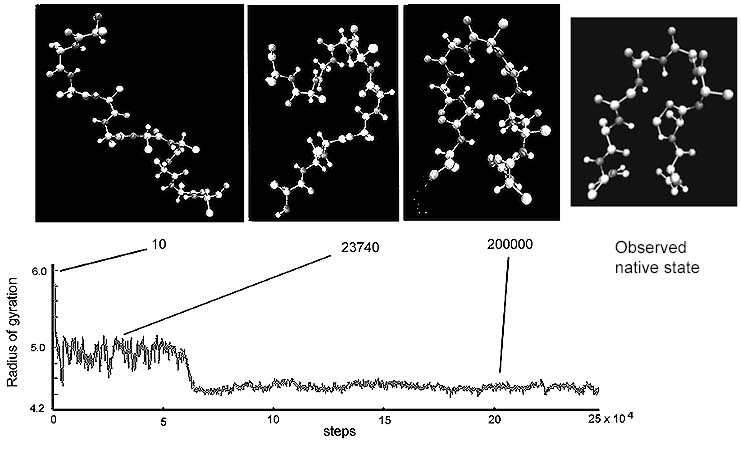}{\special{language
"Scientific Word";type "GRAPHIC";maintain-aspect-ratio TRUE;display
"USEDEF";valid_file "F";width 4.3846in;height 2.6757in;depth
0pt;original-width 7.4097in;original-height 4.51in;cropleft "0";croptop
"1";cropright "1";cropbottom "0";filename
'figs/chig_folding.jpg';file-properties "XNPEU";}}

To display the folding pathways, we project them onto a two-dimensional
subspace of the configuration space, chosen as follows. Define a $10\times
10 $ distance matrix $D_{ij}=|\mathbf{R}_{i}-\mathbf{R}_{j}|$, where $%
\mathbf{R}_{i}$ is the vector position of the $i$th $C_{\alpha }$. Let its
eignevalues be $\lambda _{1},\ldots ,\lambda _{10}$ in ascending order.
Through experimentation, we find that it is best to project the pathways
onto the $\lambda _{1}$-$\lambda _{10}$ plane, and we rotate the viewpoint
to obtain the clearest representation. This is achieved by using $\lambda
_{1}$ and $\lambda _{1}+\lambda _{10}$ as axes. Fig.9 shows the evolution of
100 folding paths. We can see that the ensemble of 100 points, identified by
given shading, migrates towards an attractor as time goes on. The energy
landscape is shown below the migration map. \FRAME{ftbpFU}{3.1929in}{4.721in%
}{0pt}{\Qcb{Evolution of 100 folding paths of Chignolin. The ensemble
evolves towards an attractor. Lower panel shows the energy landscape. See
text for explanation of the axes. }}{}{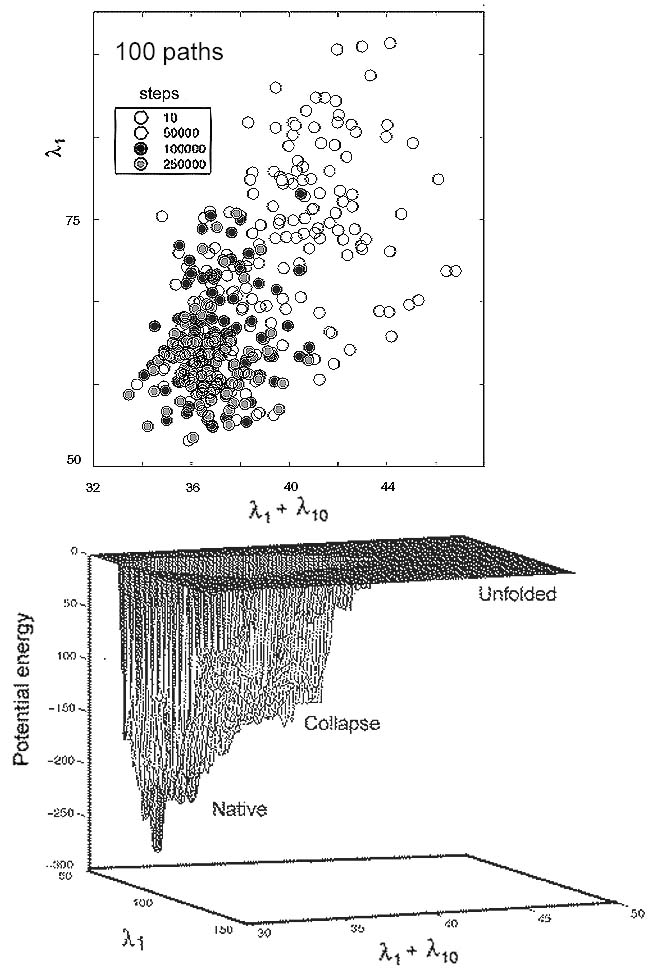}{\special{language
"Scientific Word";type "GRAPHIC";maintain-aspect-ratio TRUE;display
"USEDEF";valid_file "F";width 3.1929in;height 4.721in;depth
0pt;original-width 6.5397in;original-height 9.6997in;cropleft "0";croptop
"1";cropright "1";cropbottom "0";filename
'figs/chig_landscape.jpg';file-properties "XNPEU";}}

In Fig.10 we show 4 individual paths. They get trapped in various local
pockets, and breakout after long searches for outlets. In this respect, the
paths are similar to Levy flights.\FRAME{ftbpFU}{2.9248in}{3.3849in}{0pt}{%
\Qcb{Invidual pathways in the folding of Chignolin. Starting point are
marked with an open circle, and endpoints are marked 1.}}{}{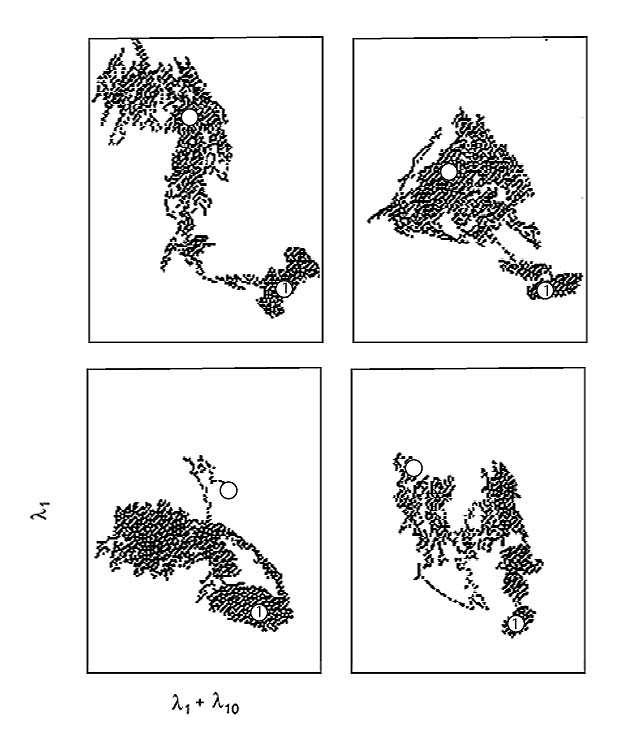}{%
\special{language "Scientific Word";type "GRAPHIC";maintain-aspect-ratio
TRUE;display "USEDEF";valid_file "F";width 2.9248in;height 3.3849in;depth
0pt;original-width 6.4403in;original-height 7.4599in;cropleft "0";croptop
"1";cropright "1";cropbottom "0";filename
'figs/chig_paths.jpg';file-properties "XNPEU";}}

Finally, in Fig.11, we exhibit the elastic property of the protein chain by
plotting the energy as a function of molecular radius, in a semilog plot.
The behavior is consistent with an exponential force law. The flat portion
in the middle corresponds to the breaking of hydrogen bonds that held the
beta hairpin together.

\FRAME{ftbpFU}{3.2448in}{2.1811in}{0pt}{\Qcb{Elastic property of Chignolin:
semilog plot of potential energy vs. radius, averaged over an ensemble of 50
samples. The flat part corresponding to the breaking of hydrogen bonds. The
general shape of the curve is consistent with an exponential force law.
Energy unit is not calibrated.}}{}{elastic1.jpg}{\special{language
"Scientific Word";type "GRAPHIC";maintain-aspect-ratio TRUE;display
"USEDEF";valid_file "F";width 3.2448in;height 2.1811in;depth
0pt;original-width 4.0603in;original-height 2.7198in;cropleft "0";croptop
"1";cropright "1";cropbottom "0";filename
'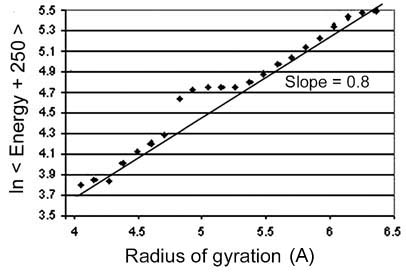';file-properties "XNPEU";}}

\section{Nucleation and growth of an alpha helix}

Next we report on Polyalanine (Ala$_{20}$), a protein fragment of 20
identical amino acids alanine, which is hydrophobic\cite{lei}. The native
state is known to be a single alpha helix. We tune $g_{1}/k_{B}T$ and $%
g_{2}/k_{B}T$ \ to maximize helical content.

An ensemble of 100 folding paths was generated. Fig.12 shows the fractions
of unfolded, intermediate, and folded molecules, as functions of time. The
solid curves are fits made according to a specific model, in which the
molecular radius reaches equilibrium first, while the helical content
continues to grow. The helical growth is described by a set of rate
equations, while the relaxation of the radius is akin to that of an elastic
solid. This shows that the tertiary structure was established before the
secondary structure, and their evolutions are governed by different
mechanisms.\FRAME{ftbpFU}{3.3356in}{3.1038in}{0pt}{\Qcb{Fractions of
Polyalanine at various stages of folding, as functions of time. Picuture at
top shows the native state of the protein fragment.}}{}{ala_rates.jpg}{%
\special{language "Scientific Word";type "GRAPHIC";maintain-aspect-ratio
TRUE;display "USEDEF";valid_file "F";width 3.3356in;height 3.1038in;depth
0pt;original-width 3.2897in;original-height 3.0597in;cropleft "0";croptop
"1";cropright "1";cropbottom "0";filename
'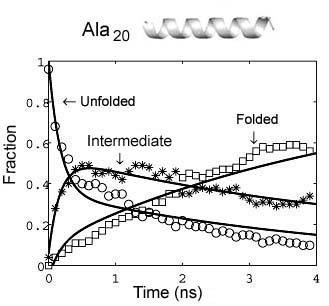';file-properties "XNPEU";}}

Fig.13 shows a contour plot of the ensemble average of helicity, with time
on the horizontal axis, and residue number along the vertical. We can see
that the alpha helix grew from two specific nucleation points.\FRAME{ftbpFU}{%
3.0588in}{2.4725in}{0pt}{\Qcb{Contour map of ensemble average of helicity as
a function of time and residue sequence, in the folding of Polyalanine.
Nuclearion occured near the two positions marked by arrows.}}{}{%
ala_nucleate.jpg}{\special{language "Scientific Word";type
"GRAPHIC";maintain-aspect-ratio TRUE;display "USEDEF";valid_file "F";width
3.0588in;height 2.4725in;depth 0pt;original-width 6.89in;original-height
5.5599in;cropleft "0";croptop "1";cropright "1";cropbottom "0";filename
'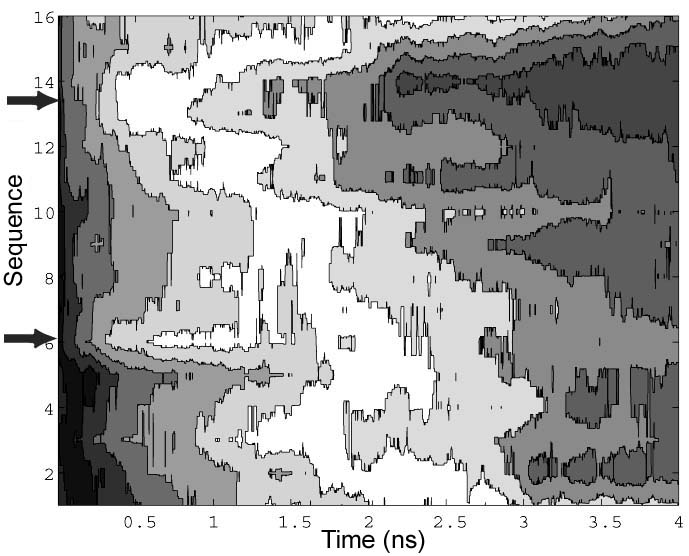';file-properties "XNPEU";}}

\section{All-atom model}

Finally we show some preliminary results of Weitao Sun \cite{sun} of
Tsinghua University on the histone 1A7W, which has 68 residues. This is a
test of an all-atom CSAW model, in which atoms on the side chains are
explicitly included. The model also includes the electrostatic interactions
among all atoms. Fig.14 compares the simulated shape of the protein with the
native state. It was found that inclusion of electrostatic interactions
makes a noticeable improvement.

\FRAME{ftbpFU}{2.7207in}{1.4339in}{0pt}{\Qcb{Folding the histone 1A7W (68
residues) with an all-atom CSAW model including electrostatic interactions.}%
}{}{histone.jpg}{\special{language "Scientific Word";type
"GRAPHIC";maintain-aspect-ratio TRUE;display "USEDEF";valid_file "F";width
2.7207in;height 1.4339in;depth 0pt;original-width 4.1597in;original-height
2.1793in;cropleft "0";croptop "1";cropright "1";cropbottom "0";filename
'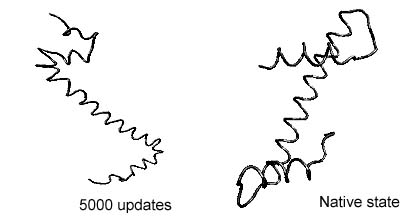';file-properties "XNPEU";}}

\FRAME{ftbpF}{3.4644in}{4.7227in}{0in}{}{}{skw.jpg}{\special{language
"Scientific Word";type "GRAPHIC";maintain-aspect-ratio TRUE;display
"USEDEF";valid_file "F";width 3.4644in;height 4.7227in;depth
0in;original-width 4.5904in;original-height 6.2699in;cropleft "0";croptop
"1";cropright "1";cropbottom "0";filename '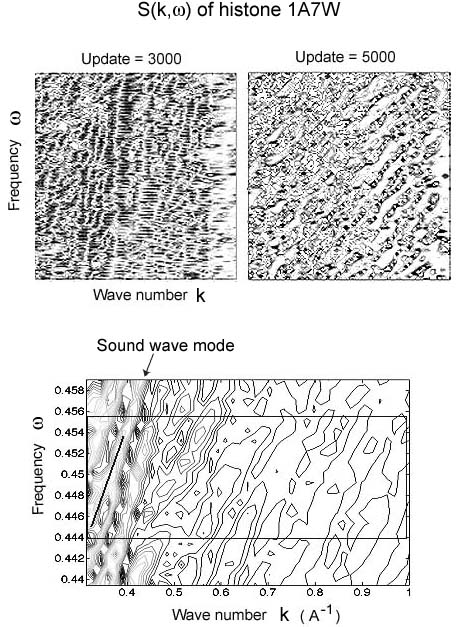';file-properties
"XNPEU";}}The main purpose of this calculation is to study the evolution of
the dynamical structure function%
\begin{equation*}
S(k,\omega )=\left\langle \left\vert n(\mathbf{k},\omega )\,\right\vert
^{2}\right\rangle 
\end{equation*}%
where $n(\mathbf{k},\omega )$ is the space-time Fourier transform of the
particle density, and $\langle \rangle $ denotes ensemble average. In
principle, this function can be experimentally measured via inelastic x-ray
scattering. A peak occurring at particular $k,\omega $ will indicate the
existence of an excitation mode. The integral of $S(k,\omega )$ over $k$
will yield the normal mode spectrum, and the integral over $\omega $ will
yield the static structure factor. Preliminary results are shown in Fig.15.
We see that at update=3000 the collective modes of the final structure have
not yet formed, but they emerge at update=5000. The lower panel of Fig.15
shows details of a sound-wave mode with constant velocity.

\section{Discussion and outlook}

In treating protein folding as a physical process, the CSAW model differs
from MD in two important aspects, namely

\begin{itemize}
\item irrelevant degrees of freedom are ignored;

\item the environment is treated as a stochastic medium.
\end{itemize}

\noindent These, together with simplifying treatment of interactions, enable
the model to produce qualitatively correct results with minimal demands on
computer time.

An important simplification is separating the hydrophobic effect and
hydrogen bonding, as expressed by the separate terms in the potential energy
(\ref{energy}). Since both effects arise physically from hydrogen bonding,
it is not obvious that we can make such a separation. The implicit
assumption is that hydrogen bonding with water involves only the side
chains, while internal hydrogen bonding involves only atoms along the main
chain. This property is supported by statistical data, but should be a
result rather an assumption of the model. We should try to remedy this in an
improved version of the model.

The successful examples discussed here deal either with the alpha helix or
the beta hairpin. Our next goal is to study the formation of a beta sheet.
This is a much more difficult problem, for it involves global instead of
local properties of the protein chain. Not knowing which elements are
crucial for the project, we have made the following enhancements to-date:

\begin{itemize}
\item All-atom side chains can now be installed, with fractional
hydrophobicity.

\item Electrostatic interactions among all atoms can be included.

\item Hard-sphere repulsions between atoms can be replaced by Lennard-Jones
potentials.

\item Hydrogen bonds can switch among qualifying partners, with given
probability.
\end{itemize}

We hope to make progress on this problem.


\begin{thebibliography}{99}
\bibitem{bra} C. Branden and J. Tooze, \textit{Introduction to Protein
Structure}, 2nd ed., (Garland Publishing, New York, 1999).

\bibitem{dag} V. Daggett and A.R. Fersht, \textit{Nat. Rev.: Mol. Cell Biol. 
} \textbf{4}, 497 (2003).

\bibitem{kan} M. Kanehisa, \textit{Post-Genome Informatics}, (Oxford
University Press, Oxford, 2000).

\bibitem{MD} H.A. Scheraga, M. Khalili, and A. Liwo, \textit{Annu. Rev.
Phys. Chem.}, \textbf{58}, 57 (2007).

\bibitem{hualec} K. Huang. \textit{Lectures on Statistical Physics and
Protein Folding} (World Scientific Publishing, Singapore, 2005).

\bibitem{flo} P. Flory, \textit{Principles of polymer chemistry} (Cornell
University Press, London, 1953).

\bibitem{hua0} K. Huang, ``CSAW: Dynamical model of protein folding",
arXiv:cond-mat/0601244 v1 12 Jan 2006.

\bibitem{hua1} K. Huang, \textit{Biophys. Rev. Lett.}, \textbf{2}, 139
(2007).

\bibitem{huastat} K. Huang. \textit{Introduction to Statistical Physics}
(Taylor \& Francis, London, 2001) Chaps.16,17.

\bibitem{li} B. Li, N. Madras, and A.D. Sokal, \textit{J. Stat. Phys.} 
\textbf{80}, 661 (1995).

\bibitem{ken} T. Kennedy, \textit{J. Stat. Phys.} \textbf{106}, 407 (2002).

\bibitem{water} M. Matsumoto, S. Saito, and I. Ohmine, \textit{Nature}, 
\textbf{416}, 409 (2002).

\bibitem{chignolin} A. Suenga \textit{et. al.} Chem. Asian J. \textbf{2},
591 (2007).

\bibitem{lei0} L.Z. Lei (unpublished).

\bibitem{lei} L.Z. Lei and K. Huang, \textquotedblleft Dynamics of
alpha-helix formation in the CSAW model\textquotedblright , arXiv 0706 3256
v1 [cond-mat.soft] 22 Jun 2007.

\bibitem{sun} W.T. Sun (unpublished).
\end{thebibliography}
\end{document}